\documentclass[referee]{cjaa}           % referee version: for submission
\usepackage{graphicx}                   %for PS/EPS graphics inclusion, new
\input{epsf.sty}                        %for PS/EPS graphics inclusion, old
\input{psfig.sty}                       %for PS/EPS graphics inclusion, old

\begin{document}

   \title{Gamma-Ray Burst Sequences in Hardness Ratio-Peak Energy Plane\footnote{Supported by the National Science Foundation of China.}
 \mailto{}}
   \volnopage{Vol.0 (200x) No.0, 000--000}
   \setcounter{page}{1}

   \author{Xiao-Hong Cui
      \inst{1,4}
%% Please move "\mailto{}" to the corresponding author of the paper.
%% For single author or all the authors from an institute, use "\inst{}" only
%% Here is an example of three authors come from different institutes.
   \and En-Wei Liang
      \inst{1,2,3}
   \and Rui-Jing Lu
      \inst{1,2,4}
      }
   \offprints{X.-H. Cui}

   \institute{National Astronomical Observatories/Yunnan Observatory,
        \\Chinese Academy of Sciences,
              Kunming 650011; ciwei8008@163.com\\
%             \email{ciwei8008@163.com}
%             \Supported{by the National Science Foundation of
 %            China.}
%% Please give the E-mail address of the author, to whom future correspondence and
%% offprint requests will be sent. Note to pair \mailto{} with \email{}
        \and
             Department of Physics, Guangxi University, Nanning 530004, China\\
%             \email{luruijing@tom.com}
        \and
             Department of Astronomy, Nanjing University, Nanjing 210093, China\\
%             \email{ewliang@nju.edu.cn}
        \and
             The Graduate School of the Chinese Academy of Sciences, Beijing,
            China\\
          }

   \date{Received~~2004 August 8; accepted~~2004~~October 13}

   \abstract{
The narrowness of the distribution of the peak energy of $\nu F_{\nu}$ spectrum of gamma-ray bursts (GRBs) and the
unification of GRB population are great puzzles yet to be solved. We investigate the two puzzles based on the global
spectral behaviors of different GRB population, the long GRBs, the short GRBs, and the X-ray flashes (XRFs), in the
$HR-E_{\rm{p}}$ plane (HR the spectral hardness ratio) with BATSE and HETE-2 observations. It is found  that the long
GRBs and the XRFs observed by HETE-2 seem to follow the same sequence in the $HR-E_{\rm{p}}$ plane, with the XRFs at
the low end of this sequence. We fit the sequence by a universal Band function, and find that this sequence is mainly
defined by the low energy index $\alpha$, and is insensitive to the high energy index, $\beta$. With fixed $\beta=-5$,
a best fit is given by $\alpha=-1.00$ with $\chi^2_{\min}/{\rm dof}=2.2$. The long and short GRBs observed by BATSE
follow significantly different sequences in the $HR-E_{\rm p}$ plane, with most of the short GRBs having a larger
hardness ratio than the long GRBs at a given $E_{\rm{p}}$. For the long GRBs, a best-fit gives $\alpha=-0.30$ and
$\beta=-2.05$. For the short GRBs, a best fit gives $\alpha=-0.60$ with $\chi^2_{\min}=1.1$ (with $\beta$ fixed at
$-2.0$ because it is numerically unstable). The $\alpha$ value for the short GRBs is significantly greater than that
for the long GRBs. These results indicate that the global spectral behaviors of the long GRB sample and the XRF sample
are similar, while that of short GRBs is different. The short GRBs seem to be a unique subclass of GRBs, and they are
not the higher energy extension of the long GRBs.
   }

   \authorrunning{Cui, Liang, \& Lu}
   \titlerunning{GRB Sequences in $HR-E_{\rm p}$ Plane}

   \maketitle

\section{Introduction}
\label{sect:intro} Gamma-Ray Bursts (GRBs) are short and intense
gamma-ray radiations from cosmological distances. Much progress on
GRBs and their afterglows has been made in the recent decade
(Fishman \& Meegan 1995; Piran 1999; van Paradijs et al. 2000;
Cheng \& Lu 2001; M\'{e}sz\'{a}ros 2002; Zhang \& M\'{e}sz\'{a}ros
2004; Piran 2004). It is believed today that GRBs are produced by
conical ejecta (jet) powered by central engines at cosmological
distances. Numerous observations of this phenomenon have been
successfully explained by the popular fireball models. However,
this phenomenon is still much mysterious. A great number of
puzzles are still to be solved.

Unification of GRB population is one of the puzzles. Based on
their distribution in the burst duration-harness ratio ($HR$)
plane, Kouveliotou et al. (1993) suggested two subclasses of GRBs,
long GRBs and short GRBs separated at $\sim 2$ seconds (see also
Qin et al. 2001). In recent years, a new subclass of GRBs, the
X-ray flashes (XRFs), was also discovered (e.g., Heise et al.
2001). Whether or not the different subclasses of GRBs are the
same phenomenon? How to provide a unified description for GRB
population? These questions have been gotten much attention (Heise
et al. 2001; Kippen et al. 2003; Sakamoto et al. 2004a, 2004b;
Lamb et al. 2003; Yamazaki et al. 2003a, 2003b, 2004a, 2004b;
Lloyd-Ronning et al. 2004; Zhang et al. 2004; Dai \& Zhang 2004;
Liang \& Dai 2004).

The peak energy of the $\nu F_{\nu}$ spectrum, $E_{\rm{p}}$, is an
important quantity of GRB. Liang et al. (2002a) showed a
limitation of $E_{\rm{p}}$ with burst duration. This limitation
cannot be explained by the current fireball model. It may
represent a constraint on the fireball model. The narrowness of
the $E_{\rm{p}}$ distribution is also a great puzzle. The
distribution based on a time-resolved spectral catalog of bright
BATSE GRBs (Preece et al. 2000) shows that $E_{\rm{p}}$ is
narrowly clustered at 200-400 keV. It might be caused by the
selection effect of the BATSE, which is sensitive to photons in
energy band 50-300 keV. HETE-2 is sensitive to photons in an
energy band of $\sim 10-400$ keV, and it is suitable for
observations to XRFs. It is found that the spectra of XRFs are
well fitted by the Band function (Band et al. 1993) with similar
spectral indices as long GRBs (e.g., Kippen et al. 2003; Barraud
et al. 2003), and they obey the same relations of $E_{\rm{p}}$ to
the equivalent-isotropic energy (Amati et al. 2002; Lloyd-Ronning
\& Ramirez-Ruiz 2002; Sakamoto et al. 2004a; Lamb et al. 2003;
Liang, Dai, \& Wu 2004; Yonetoku et al. 2004) and to the jet
energy (Ghirlanda, Ghisellini \& Lazzati 2004b; Dai, Liang, \& Xu
2004). The duration distributions of long GRBs and XRFs are also
similar (Heise et al. 2001). These facts suggest that long GRBs
and XRFs are the same phenomenon, and XRFs are a lower peak-energy
extension of long GRBs (e.g., Kippen et al. 2003; Sakamoto et al.
2004a). Therefore, XRFs extend the $E_{\rm{p}}$ distribution to
few keVs. Whether or not the $E_{\rm{p}}$ distribution can be
broadened to a higher energy is still uncertain (e.g., Piran
2004). Short GRBs tend to have harder spectra. Are short GRBs to
be a higher energy extension of long GRBs? Ghirlanda, Ghisellini
\& Celotti (2004a) presented an analysis to short GRBs. They found
that the spectra of short bursts are well fitted by a single power
law with an exponential cutoff at high energies. The statistics in
the high energy channels of the spectra is too poor to constrain
the high energy power law component of the Band function. They
also found that spectral properties of short GRBs are similar to
the first 1 second of long GRBs. Yamazaki et al. (2004b) proposed
a unified description for long GRBs, short GRBs, and XRFs based on
an off-axis jet model. If long and short GRBs are the same
phenomenon, one may expect that short GRBs are a higher energy
extension of long GRBs, and long GRBs, short GRBs, and XRFs form a
sequence in the $HR-E_{\rm{p}}$ plane, since the short GRBs tend
to have a harder spectrum. In this work we investigate this issue
with the current GRB samples observed by BATSE and HETE-2.

This paper is arranged as follows. The method of analysis is
presented in Section 2. The global spectral behaviors of the long
GRB and XRF samples observed by HETE-2 are shown in Section 3, and
those of the long and short GRB samples observed by BATSE, in
Section 4. A discussion and conclusions are presented in Section
5.

\section{Analysis Method}
\label{sect:analysis} It is well known that GRB spectrum can be well fitted by the Band function
(Band et al. 1993),
\begin{equation}\label{}
N(E) = A\left \{\begin{array}{l}
(\frac{E}{100keV})^{\alpha}\exp(-E/E_{0})\ \ \ \ \ \ \ \ \ \ \ \ \ \ \ \ \ \ \ \ E\leq(\alpha-\beta)E_{0},\\
(\frac{(\alpha-\beta)E_{0}}{100keV})^{\alpha-\beta}\exp(\beta-\alpha)(\frac{E}{100keV})^{\beta}
    \ \ \ \ \ E\geq(\alpha-\beta)E_{0},
\end{array} \right.
\end{equation}
where $A$ is a normalization parameter, $E_{0}$ the break energy
(in keV), $\alpha$ and $\beta$ the low and high band spectral
indices, respectively. For the case of $\beta<-2$ and $\alpha>-2$,
the peak energy can be derived by $E_{\rm{p}}=(2+\alpha)E_{0}$,
which corresponds to the energy at the maximum flux in $\nu f_\nu$
spectrum. For a spectrum with $\alpha>-1$ and $\beta>-2$, the
``true" $E_{\rm p}$ lies at an unknown energy beyond the high end
of the data, and only a lower boundary, $E_{\rm{b}}$, to the
energy of the high-energy power-law component characterized by
$\beta$ is known. The observed fluence at a given bandpass ($E_1$,
$E_2$) as a function of $E_{\rm{p}}$ can be given by
\begin{equation}\label{}
 S_{\rm{E_1-E_2}}(E_{\rm{p}})=\int^{E_2}_{E_1}EN(E)dE.
\end{equation}

The hardness ratio is defined as

\begin{equation}\label{}
HR(E_{\rm{p}})\equiv\frac{S_{\rm{E_1-E_2}}(E_{\rm{p}})}{S_{\rm{E_3-E_4}}(E_{\rm{p}})},
\end{equation}
and its error caused by $E_{\rm{p}}$ is (without consider the
errors of $\alpha$ and $\beta$ in our analysis),

\begin{equation}
\sigma_{HR}=HR\sqrt{(\frac{\sigma_{S_{\rm{E_3-E_4}}}}{S_{\rm{E_3-E_4}}})^2+(\frac{\sigma_{S_{\rm{E_1-
E_2}}}}{S_{\rm{E_1-E_2}}})^2},
\end{equation}
where

\begin{equation}\label{}
\sigma_{S_{\rm{E_i-E_j}}}=\left \{\begin{array}{l}
\frac{|2+\alpha|}{E_{\rm{p}}} \frac{\sigma_{E_{\rm{p}}}}{E_{\rm{p}}}\int^{E_{\rm{j}}}_{E_{\rm{i}}} E^2 NdE\ \ \ \ \ \ \ \ \ \ \ \ \ \ \ \ \ \ \ \ E\leq(\alpha-\beta)E_{0},\\
\frac{|\alpha-\beta|\sigma_{E_{\rm{p}}}}{E_{\rm{p}}}\int^{E_{\rm{j}}}_{E_{\rm{i}}}EN(E)dE
\ \ \ \ \ \ \ \ \ \ \ \ \ \ \ \ \ E\geq(\alpha-\beta)E_{0}.
\end{array} \right.
\end{equation}

GRBs with similar spectral behaviors should trace out a sequence
in the $HR-E_{\rm{p}}$ plane. In this paper we investigate whether
or not different subclasses of GRBs are in a $HR-E_{\rm{p}}$
sequence characterized by a universal Band function by using the
GRBs observed by BATSE and HETE-2.

\section{Long GRBs vs. XRFs}
\label{sect:results} HETE-2 is sensitive to photons in the energy
band of $\sim 10-400$ keV. It is suitable for observing GRBs and
XRFs. We examine the sequence of long GRBs and XRFs in the
$HR-E_{\rm{p}}$ plane with the bursts observed by HETE-2. Of the
63 HETE-2 bursts (to the end of June, 2004), 34 bursts are long
GRBs and 22 bursts are XRFs. We include only the long GRBs and the
XRFs in our analysis. The values of $E_{\rm{p}}$, fluences
($S_{30-400{\rm keV}}$ and $S_{2-30{\rm keV}}$, in energy bands of
$30-400$ keV and $2-30$ keV, respectively), and their errors for
40 bursts are taken from Sakamoto et al. (2004b). The other bursts
are taken from HETE-2 burst home
page\footnote{http://space.mit.edu/HETE/Bursts/} and G. Ricker's
report (private communication), but no errors of $E_{\rm{p}}$ and
fluences are available for these. The errors of GRB030324 and
GRB030723 given by Sakamoto et al. (2004b) are extremely large. We
calculate the average percentages of the errors for the bursts
from Sakamoto et al. (2004b) not includings GRB030324 and
GRB030723, and obtain $<\sigma_{E_{\rm{p}}}/E_{\rm{p}}>\simeq
0.19$, $<\sigma_{S_{30-400 {\rm keV}}}/S_{30-400{\rm keV}}>\simeq
0.26$, $<\sigma_{S_{2-30{\rm keV}}}/S_{2-30{\rm keV}}>\simeq
0.09$. Thus, for those bursts with no errors available, and for
the two bursts, GRB030324 and GRB030723, we take
$\sigma_{E_{\rm{p}}}=0.19 E_{\rm{p}}$, $\sigma_{S_{30-400 {\rm
keV}}}=0.26S_{30-400{\rm keV}}$, $S_{2-30{\rm keV}} \simeq
0.09S_{2-30{\rm keV}}$. The $HR^{ob}$ is calculated by
$HR^{ob}=S_{30-400{\rm keV}}/S_{2-30{\rm keV}}$, and its error is
given by Eq. (4) with measured errors of the fluences in the two
energy bands. $HR^{ob}$ as a function of $E_{\rm{p}}$ for the
HETE-2 bursts is shown in Fig. 1. From Fig. 1, we find that the
GRBs and the XRFs seem to form a well-defined sequence. We fit
this sequence with Eq. (3) by minimizing the $\chi^2$,

\begin{equation}
\chi^2=\sum_i \frac{(HR_{i}^{ob}-HR_{i}^{th})^2}{\sigma_{HR^{ob}_i}^2+\sigma_{HR^{th}_i}^2}
\end{equation}
where $\sigma_{HR_{i}^{th}}$ is the error of $HR$ caused by the
uncertainty of $E_{\rm{p}}$, which is calculated by Eqs. (4) and
(5). In our calculations, we find that this sequence is
characterized by $\alpha$; while $\beta$ is numerically unstable
(varying it from $-5$ to $-10$ gives comparable values of
$\chi^2_{\rm{\min}}$ ($\sim 118$)). So we fix $\beta=-5$, and then
make a best fit to the sequence, and obtain $\alpha=-1.00$ with
$\chi^2_{\min}/{\rm dof}=2.2$. The best fit curve is also plotted
in Fig. 1 (the solid curve). These results show that both the GRBs
and the XRFs can be described by a universal Band function with
$E_{\rm{p}}$ ranging from few keVs to hundreds of keVs. The XRFs
are at the lower end of the long GRB sequence, suggesting that
XRFs are the lower energy extension of the long GRB sequence.

\begin{figure}[t]
\vbox to 2.0in{\rule{0pt}{2.0in}} \includegraphics{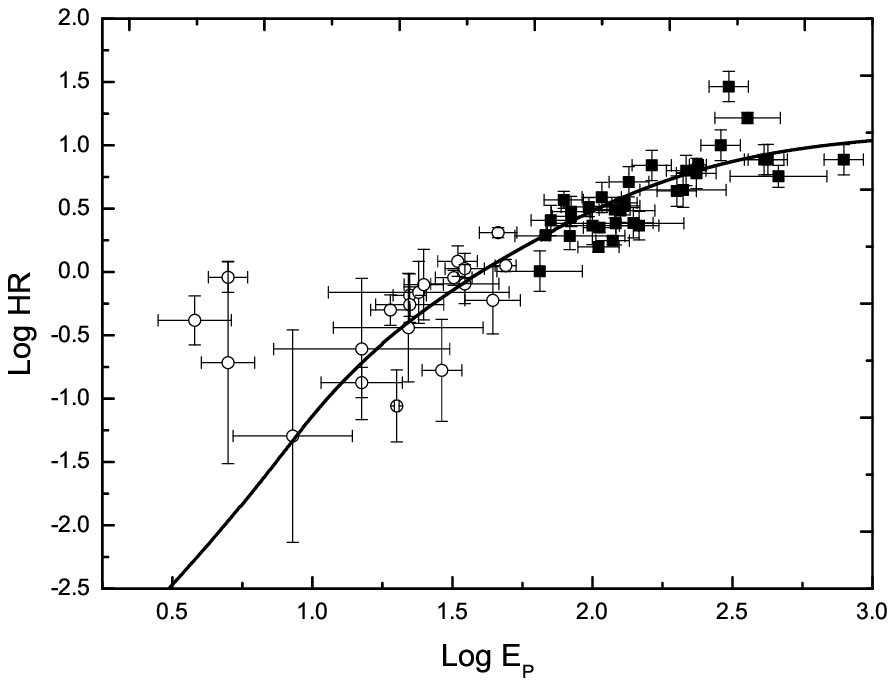}
\caption{$HR$ as a function of $E_{\rm{p}}$ for the long GRBs (solid circles) and XRFs (open circles) observed by
HETE-2. The solid curve is the best fit with Eq. (3) by fixing $\beta=-5$.}
\end{figure}

\section{Long GRBs vs. Short GRBs}

In the present GRB sample, most of the short GRBs were observed by
BATSE. To investigate whether or not both short and long GRBs can
be characterized by a universal Band function, we use the long
BATSE GRB sample (149 GRBs) given by Lloyd-Ronning \& Ramirez-Ruiz
(2002) and the short BATSE GRB sample (28 GRBs) presented by
Ghirlanda et al. (2004a). The fluences and their errors are
available at the FLUX TABLE of the Current BATSE Catalog
\footnote{http://cossc.gsfc.nasa.gov/batse/}. We calculate the
hardness ratios by the observed fluences in channel 3 ($110-320$
keV) to those of channel 2 ($55-110$ keV), and their errors are
derived by Eq. (4) with the errors of the fluences in channels 2
and 3. The $HR$ versus $E_{\rm{p}}$ plot for the long BATSE GRB
sample is shown in Fig. 2. Since no $E_{\rm{p}}$ errors of these
bursts are available in Lloyd-Ronning \& Ramirez-Ruiz (2002), we
fit the long GRB sequence by Eq. (3) without considering the
errors. The result shows that the minimum residual square,
$\mu^2=\sum_i(HR^{ob}_i-HR^{th}_{i})^2$, is 131 at ($\alpha$,
$\beta$)=(-0.3, -2.05), which is also shown in Fig. 2.

The $HR$ versus $E_{\rm{p}}$ plot for the short BATSE GRB sample
is shown in Fig. 3. The spectrum of these short bursts is well
fitted by a single power law with an exponential cutoff at high
energies. Besides, the statistics in the high energy channels of
the spectra is too poor to constrain the high energy power law
component of the Band model (Ghirlanda et al. 2004a). We make a
best fit to the short GRB sequence with Eq. (3). We find that
$\beta$ is ill-defined. So we fix $\beta=-2$ (similar to the long
BATSE GRBs), and then make a best fit to the sequence, and derives
$\alpha=-0.60$ with $\chi^2_{\min}/{\rm dof}=1.1$. The $\alpha$
value is quite similar to the mean of $\alpha$ in Ghirlanda et al.
(2004a) (-0.58). The best fit curve is plotted in Fig. 3 (the
solid curve). For comparison, the best fit for the long BATSE GRB
sample is also plotted in Fig. 3 (dotted curve). It is found that
the two best fits are quite different. Most of the short GRBs are
above the long GRB sequence, indicating that the short GRBs tend
to have a larger $HR$ than the long GRBs at a given $E_{\rm{p}}$.
Short GRBs are unlikely to be a higher energy extension of long
GRBs.

\begin{figure}[t]
\vbox to 2.0in{\rule{0pt}{2.0in}} \includegraphics{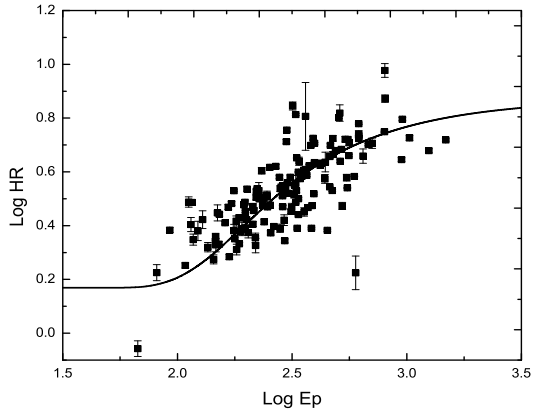}
\caption{$HR$ as a function of $E_{\rm{p}}$ for the long BATSE GRBs. The solid curve is the best fit with Eq. (3).}

\end{figure}

\begin{figure}[t]
\vbox to 2.0in{\rule{0pt}{2.0in}}

\includegraphics{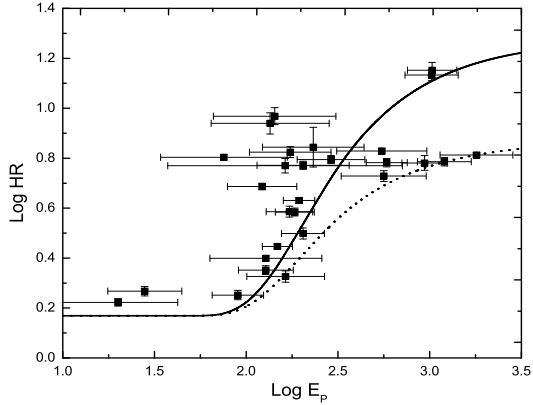} \caption{$HR$ as a function of
$E_{\rm{p}}$ for the short BATSE GRBs. The solid curve is the best with Eq. (3) with fixed $\beta=-2.0$ , and the
dotted curve is the best fit for the long BATSE GRB sample.}

\end{figure}

\section{Conclusions and discussion}
\label{sect:discussion} We have investigated the global spectral
behaviors of long GRBs, short GRBs, and XRFs in the
$HR-E_{\rm{p}}$ plane with BATSE and HETE-2 observations. It is
found that the long GRBs and the XRFs observed by HETE-2 seem to
follow the same sequence in the $HR-E_{\rm{p}}$ plane, with the
XRFs at the low end of this sequence. We fit the sequence by a
universal Band function, and find that this sequence is mainly
fixed by $\alpha$, while $\beta$ is numerically unstable, with
$\chi^2_{\rm{\min}}$ remaining at $\sim 118$ for $\beta$ varying
form $-5$ to $-10$. So we fix $\beta=-5$, and then make a best fit
to the sequence, and derive $\alpha=-1.00$ with
$\chi^2_{\min}/{\rm dof}=2.2$. The sequences of the long and short
GRBs observed by BATSE in the $HR-E_{\rm p}$ plane are
significantly different from that of the long GRBs. In the
$HR-E_{\rm{p}}$ plane, most of the short GRBs are above the long
GRB sequence, indicating that the short GRBs tend to have a larger
$HR$ than the long GRBs at a given $E_{\rm{p}}$. A best-fit gives
$\alpha=-0.30$ and $\beta=-2.05$ for the long BATSE GRBs. For the
short GRBs, a best fit is obtained with $\alpha=-0.60$, with
$\chi^2_{\min}=1.1$ ($\beta$ is fixed at $-2.0$ since it is again
numerically unstable). The $\alpha$ value of the short GRB
sequence is significantly greater than that of the long GRBs.
These results suggest strongly that the global spectral behaviors
of the long GRBs and the XRFs are similar, and the XRFs are the
lower energy extension of the long GRBs, while the global spectral
behaviors of the short GRBs are different from the long GRBs: the
short GRBs seem to be a unique subclass of GRBs, and they are not
the higher energy extension of the long GRBs.

The unified description of long GRBs and XRFs has been widely
discussed. Zhang et al. (2004) showed that current GRB/XRF prompt
emission/afterglow data can be described by a quasi-universal
Gaussian-like (or similar structure) structured jet with a typical
opening angle of $\sim 6^\circ$ and with a standard jet energy of
$\sim 10^{51}$ ergs. Based on HETE-2 observations, Lamb et al.
(2003) proposed that the uniform jet model nicely describes the
prompt emission data of GRBs/XRFs, but this model fails to account
for the afterglow jet break time data of GRBs. Liang \& Dai (2004)
found a bimodal distribution of the observed $E_{\rm{p}}$ of
GRBs/XRFs and suggested that two-component jet model (Berger et
al. 2003) can explain this distribution, proposed that the
two-component jet seems to be universal for GRBs/XRFs. The results
of this paper further confirm the view that long GRBs and XRFs are
the same phenomenon.

We show that short GRBs are not the higher energy extension of GRBs, although they tend to have
a harder spectrum than long GRBs. Please note that this result is based on the BATSE
observations. Whether or not the $E_{\rm{p}}$ distribution can be broadened to a higher energy
is still uncertain. This possibility cannot be ruled out by BATSE observations.

The observed association of long GRBs with star formation regions
(e.g., Djorgovski et al. 2001), and the possible supernova
components in afterglow light curves (e.g., Bloom et al. 1999;
Rechart et al. 2001; Stanek et al. 2003; Hjorth et al. 2003)
indicate that the central engines of the long GRBs might be the
collapses of supermassive stars to black holes (e.g., Wooseley
1993; Berger, Kulkarni, \& Frail 2003). Our results suggest that
short GRBs are not the same phenomenon as the long GRBs and the
XRFs. So far no afterglows of short GRBs have been observed. It is
not clear whether this is an observational artifact or a real
feature. Furthermore, we do not know of any direct evidence
relating to the origin of short GRBs. The results in this paper
might hint that short and long GRBs might come from different
progenitors. Liang et all. (2002b, c) made the same argument based
on the differences of the variability time scales and the total
fluences of long and short GRBs.

From Fig. 1, it is found that the sequences in the $HR-E_{\rm{p}}$
plane are mainly characterized by the value of $\alpha$,
indicating that the GRB spectra in our analysis are dominated by
photons with energies lower than $E_{\rm{p}}$. In our
calculations, we find that $\chi^2_{\min}$ is sensitive to
$\alpha$, but not to $\beta$. The $\beta$ value significantly
affects the low end of the sequence at $\log E_{\rm{p}}/{\rm
keV}<1$. However, there are only four XRFs in our GRB sample in
this range. Thus, we cannot well constrain the low tail of the GRB
sequence (and then the value of $\beta$). In fact, the $\beta$
value is ill-defined. This indicates that the statistics in the
high energy channels of the spectra is too poor to constrain the
high energy power law component of the Band model (Ghirlanda et
al. 2004a).

It should be  noted that our best fit to the GRB sequence in the
$HR-E_{\rm{p}}$ plane is related to the definition of $HR$.
Different $HR$ definition can lead to different best fit results
for the same GRB sample. This is the reason why the best fits
shown in Fig. 1 and 2 for HETE-2 long GRB sample and for BATSE
long GRB sample are quite different. For the BATSE GRB sample,
$HR$ is calculated by the fluences in Channel 3 and 2(energy bands
110-320 keV and 55-110 keV) while the $HR$ of the HETE-2 GRB
sample is calculated by the fluences in the energy bands 30-400
keV and in 2-30 keV. Therefore, the best-fits to the HETE-2 GRB
sample and to the BATSE GRB sample are different. However, this
difference does not affect the results of our analysis because we
separately investigated the long GRB-XRF sequence and the long
GRB-short GRB sequence by the two instruments.

\begin{acknowledgements}
We are very grateful to the anonymous referee for his/her valuable
suggestions. We thank Dai Z. G, Qin Y. P., Wei D. M., and Wu X. F.
for their helpful comments. These suggestions and comments have
enabled us to improve greatly the manuscript. We also thank Dr. G.
Richer to provide some unpublished data. This work is supported by
the National Natural Science Foundation of China (No. 10463001 and
No.10273019), the Postdoctoral Foundation of China (No.
20040350213), the National ¡°973¡± Project, and the Research
Foundation of Guangxi University.
\end{acknowledgements}

\label{lastpage}

\end{document}